\documentclass[a4paper]{amsart}
\usepackage{epic,eepic,amsmath,amsthm,amssymb}
\usepackage{epsfig,cite,indentfirst,oldgerm}
\usepackage{multicol,boxedminipage}

\begin{document}

\title[BKP Plane Partitions]
{BKP Plane Partitions}

\author{O Foda and M Wheeler}

\address{Department of Mathematics and Statistics,
         University of Melbourne, 
         Parkville, Victoria 3010, Australia.}
\email{foda@ms.unimelb.edu.au, mwheeler@ms.unimelb.edu.au}

\date{}

\newcommand{\field}[1]{\mathbb{#1}}
\newcommand{\C}{\field{C}}
\newcommand{\N}{\field{N}}
\newcommand{\Z}{\field{Z}}

\begin{abstract}
Using BKP neutral fermions, we derive a product expression for the 
generating function of volume-weighted plane partitions that satisfy 
two conditions. If we call a set of adjacent equal height-$h$ columns, 
$h > 0$, an $h$-path, then {\bf 1.} Every $h$-path can assume one of 
two possible colours. {\bf 2.} There is a unique way to move along 
an $h$-path from any column to another. 
\end{abstract}

\maketitle
\newtheorem{ca}{Figure}
\newtheorem{corollary}{Corollary}
\newtheorem{de}{Definition}
\newtheorem{definition}{Definition}
\newtheorem{example}{Example}
\newtheorem{ex}{Example}
\newtheorem{lemma}{Lemma}
\newtheorem{no}{Notation}
\newtheorem{proposition}{Proposition}
\newtheorem{pr}{Proposition}
\newtheorem{remark}{Remark}
\newtheorem{re}{Remark}
\newtheorem{theorem}{Theorem}
\newtheorem{theo}{Theorem}

\def\ll{\left\lgroup}
\def\rr{\right\rgroup}

\newcommand{\Proof}{\medskip\noindent {\it Proof: }}
\def\no{\nonumber}
\def\proofend{\ensuremath{\square}}
\def\Gamc{\Gamma^{C}}
\def\pr{'}
\def\beqa{\begin{eqnarray}}
\def\eeqa{\end{eqnarray}}
\def\ba{\begin{array}}
\def\ea{\end{array}}
\def\gl{\begin{swabfamily}gl\end{swabfamily}}
\def\psis{\psi^{*}}
\def\Psis{\Psi^{*}}
\def\union{\mathop{\bigcup}}
\def\vac{|\mbox{vac}\rangle}
\def\cav{\langle\mbox{vac}|}
\def\lprod{\mathop{\prod{\mkern-29.5mu}{\mathbf\longleftarrow}}}
\def\rprod{\mathop{\prod{\mkern-28.0mu}{\mathbf\longrightarrow}}}
\def\r{\rangle}
\def\l{\langle}
\def\a{\alpha}
\def\b{\beta}
\def\hb{\hat\beta}
\def\d{\delta}
\def\g{\gamma}
\def\e{\epsilon}
\def\tg{\operatorname{tg}}
\def\ctg{\operatorname{ctg}}
\def\sh{\operatorname{sh}}
\def\ch{\operatorname{ch}}
\def\cth{\operatorname{cth}}
\def\th{\operatorname{th}}
\def\eps{\varepsilon}
\def\la{\lambda}
\def\tla{\tilde{\lambda}}
\def\tmu{\tilde{\mu}}
\def\s{\sigma}
\def\sul{\sum\limits}
\def\pl{\prod\limits}
\def\lt({\left(}
\def\rt){\right)}
\def\pd #1{\frac{\partial}{\partial #1}}
\def\const{{\rm const}}
\def\argum{\{\mu_j\},\{\la_k\}} 
\def\umarg{\{\la_k\},\{\mu_j\}} 
\def\prodmu #1{\prod\limits_{j #1 k} \sinh(\mu_k-\mu_j)}
\def\prodla #1{\prod\limits_{j #1 k} \sinh(\lambda_k-\lambda_j)}
\newcommand{\bl}[1]{\makebox[#1em]{}}
\def\tr{\operatorname{tr}}
\def\Res{\operatorname{Res}}
\def\det{\operatorname{det}}

\newcommand{\boldN}{\boldsymbol{N}}
\newcommand{\bra}[1]{\langle\,#1\,|}
\newcommand{\ket}[1]{|\,#1\,\rangle}
\newcommand{\bracket}[1]{\langle\,#1\,\rangle}
\newcommand{\infinity}{\infty}

\renewcommand{\labelenumi}{\S\theenumi.}

\let\up=\uparrow
\let\down=\downarrow
\let\tend=\rightarrow
\hyphenation{boson-ic
             ferm-ion-ic
             para-ferm-ion-ic
             two-dim-ension-al
             two-dim-ension-al
             rep-resent-ative
             par-tition}

\newtheorem{Theorem}{Theorem}[section]
\newtheorem{Corollary}[Theorem]{Corollary}
\newtheorem{Proposition}[Theorem]{Proposition}
\newtheorem{Conjecture}[Theorem]{Conjecture}
\newtheorem{Lemma}[Theorem]{Lemma}
\newtheorem{Example}[Theorem]{Example}
\newtheorem{Note}[Theorem]{Note}
\newtheorem{Definition}[Theorem]{Definition}
                                                                               
\renewcommand{\mod}{\textup{mod}\,}
\newcommand{\wt}{\text{wt}\,}

\newcommand{\T}{{\mathcal T}}
\newcommand{\U}{{\mathcal U}}
\newcommand{\tT}{\tilde{\mathcal T}}
\newcommand{\tU}{\widetilde{\mathcal U}}
\newcommand{\Y}{{\mathcal Y}}
\newcommand{\B}{{\mathcal B}}
\newcommand{\D}{{\mathcal D}}
\newcommand{\M}{{\mathcal M}}
\renewcommand{\P}{{\mathcal P}}
\newcommand{\R}{{\mathcal R}}

\hyphenation{And-rews
             Gor-don
             boson-ic
             ferm-ion-ic
             para-ferm-ion-ic
             two-dim-ension-al
             two-dim-ension-al}

\setcounter{section}{-1}

\section{Introduction}\label{introduction}

In \cite{OR}, Okounkov and Reshetikhin observed that certain charged 
free fermion vertex operators can be used to generate plane partitions. 
In \cite{ORV}, with Vafa, they used this observation to compute the 
partition function of a topological string theory. As these vertex 
operators arise in KP theory \cite{blue-book}, it is natural to look 
for analogous results in the context of other integrable hierarchies 
\cite{JM}. 

\begin{de} An `$h$-path' in a plane partition, or simply a `path' when 
indicating the height $h$ is not needed, is a set of adjacent equal 
height-$h$ columns, where $h > 0$.
\end{de}

In this note, we use BKP neutral free fermion vertex operators to 
obtain the generating function of volume-weighted plane partitions 
\cite{macdonald}, that satisfy two conditions. {\bf 1.} Every $h$-path, 
$h > 0$, can assume one of two possible colours, so it contributes 
a factor of 2 to the multiplicity of the plane partition, irrespective 
of $h$. {\bf 2.} There is a unique way to move along an $h$-path, from 
one column to another, or equivalently {\it `every $h$-path is 1-column 
wide'}.

\begin{de} A `BKP plane partition' is a plane partition that satisfies 
the above two conditions. 
\end{de}

\begin{example}\label{example} 
In Figure {\bf 1}, we use tableau-like notation to represent a plane 
partition of the type counted in this note. The integers are the column 
heights. The volume of a plane partition is the sum of all column 
heights. The volume in this case is 39. There are 6 $h$-paths. Each 
path can assume one of two possible colours, so the multiplicity of 
this plane partition is $2^6$ $=$ $64$. 
\end{example}

\begin{center}
\begin{minipage}{12cm}
\setlength{\unitlength}{0.001cm}

\begin{picture}(5500, 3000)(-5000, 0300)


\path(0000, 2400)(2400, 2400)
\path(0000, 1200)(2400, 1200)
\path(0000, 1800)(2400, 1800)
\path(0000, 0600)(1800, 0600)


\path(0000, 2400)(0000, 0600)
\path(0600, 2400)(0600, 0600)
\path(1200, 2400)(1200, 0600)
\path(1800, 2400)(1800, 0600)
\path(2400, 2400)(2400, 1200)


\put(2100, 2000){\makebox(0, 0)[1b] 2}
\put(2100, 1400){\makebox(0, 0)[1b] 1}
\put(2100, 0800){\makebox(0, 0)[1b]  }

\put(1500, 2000){\makebox(0, 0)[1b] 3}
\put(1500, 1400){\makebox(0, 0)[1b] 3}
\put(1500, 0800){\makebox(0, 0)[1b] 3}

\put(0900, 2000){\makebox(0, 0)[1b] 6}
\put(0900, 1400){\makebox(0, 0)[1b] 4}
\put(0900, 0800){\makebox(0, 0)[1b] 3}

\put(0300, 2000){\makebox(0, 0)[1b] 6}
\put(0300, 1400){\makebox(0, 0)[1b] 5}
\put(0300, 0800){\makebox(0, 0)[1b] 3}

\end{picture}
\begin{ca} 
\label{figure}
A tableau-like representation of a plane partition. There is 
a 3-path of length 5, 
a 6-path of length 2, and
4 different height paths of length 1 each.
There is a unique way to move from any column on a path to
another column. Counting a no-move on a length-1 $h$-path as 
the one (and only) possible move, every $h$-path is 1-column 
wide, so it qualifies as a BKP partition.
\end{ca}
\end{minipage}
\end{center}
\bigskip

\section{BKP fermions}\label{bkp}

In this section, we review basic facts related to BKP neutral 
fermions \cite{JM}.

\subsection{Neutral fermions} 

Following \cite{JM}, we consider the neutral fermion field
$\Phi(k)$ $=$ $\sum_{m \in \Z} \phi_m k^m$, where the mode 
operators, $\phi_m$, satisfy the anti-commutation relation

\begin{equation}
\left[ \phi_m, \phi_n \right]_{+} = (-)^m \delta_{m + n, 0}, 
\quad m, n \in \Z
\end{equation}

\subsection{Fock states} We indicate an initial Fock state by 
$\langle \ldots, i_2, i_1 |$, with $\ldots < i_2 < i_1 \leq 0$, 
and a final state by
$| j_1, j_2, \ldots \rangle$, with $0 \leq j_1 < j_2 < \ldots$ 
where, as usual, the integers $\{i_m, j_n\}$ indicate filled 
neutral fermion energy states. 

\subsubsection{} The action of $\phi_{m}$, $m > 0$, is 

\begin{eqnarray*}
\langle \ldots,i_2,i_1|\phi_{(m>0)} 
&=& 
\left\{
\begin{array}{ll}
(-)^{m+k} \langle \ldots, i_{k+1}, -m, i_{k}, \ldots, i_1|, & 
i_{k+1}<-m<i_{k} 
\\ 
0, & \mbox{\it otherwise} 
\end{array}
\right.
\\ \\
\phi_{(m>0)}|j_1, j_2, \ldots \rangle 
&=& 
\left\{
\begin{array}{ll}
(-)^{m+k-1}| j_1, \ldots, j_{k-1}, j_{k+1}, \ldots \rangle, &  
m = j_{k} \\ 
0, & \mbox{\it otherwise} 
\end{array}
\right.
\end{eqnarray*}

\subsubsection{} The action of $\phi_{m}$, $m < 0$, is 

\begin{eqnarray*}
\langle \ldots,i_2,i_1|\phi_{(m<0)} 
&=& 
\left\{
\begin{array}{ll}
(-)^{k-1} \langle \ldots, i_{k+1}, i_{k-1}, \ldots, i_1|, & m=i_k 
\\ 0, & \mbox{\it otherwise} 
\end{array}
\right.
\\
\phi_{(m<0)}| j_1, j_2, \ldots \rangle 
&=& 
\left\{
\begin{array}{ll}
(-)^{k}|j_1, \ldots, j_{k}, -m, j_{k+1}, \ldots \rangle, & 
j_{k}<-m<j_{k+1} 
\\ 0, & \mbox{\it otherwise} 
\end{array}
\right. 
\end{eqnarray*}

\subsubsection{} The action of {\bf $\phi_0$} is 

\begin{eqnarray*}
\langle \ldots, i_2, i_1| \phi_0 
&=& 
\left\{
\begin{array}{ll}
\frac{1}{\sqrt{2}}\langle\ldots,i_2,i_1,0|, & i_1 \not= 0 \\
\frac{1}{\sqrt{2}}\langle\ldots,i_2|, & i_1=0
\end{array}
\right.
\\
\phi_0| j_1, j_2, \ldots \rangle 
&=& \left\{
\begin{array}{ll}
\frac{1}{\sqrt{2}}|0, j_1, j_2, \ldots \rangle, & j_1 \not= 0 
\\
\frac{1}{\sqrt{2}}| j_2, \ldots \rangle, & j_1 = 0
\end{array}
\right. 
\end{eqnarray*}

\subsubsection{Remark}\label{even} Notice that $0$ is an allowed
filling number, which can be added or removed by the action of
$\phi_0$. In the following, this action is used to represent any 
Fock state in terms of an {\it even} number of mode operators 
acting on the vacuum. 

\subsection{Strict partitions} 

\begin{de} A strict partition, $\widehat{\mu}$, is a partition 
that has only distinct parts. In this note, we take the number 
of parts to be always even by allowing for at most one part of 
length 0, which agrees with Remark {\bf \ref{even}}.
\label{strict}
\end{de} 

A neutral fermion initial, or final Fock state can be labeled 
by a strict partition 

\begin{eqnarray}
\l \widehat{\mu}| &=& \alpha(-)^{r+| \widehat{\mu}|}
\l 0|
\phi_{-m_{2r}}
\ldots
\phi_{-m_1} = \alpha(-)^{r+| \widehat{\mu}|} 
\l 0|
\lprod_{j=1}^{2r}\phi_{-m_j}
\no \\
|\widehat{\mu}\r 
&=&
\alpha(-)^{r} \phi_{m_1} 
\ldots 
\phi_{m_{2r}}|0 \r = \alpha(-)^{r}
\rprod_{j = 1}^{2r} \phi_{m_j} |0 \r
\label{by}
\end{eqnarray}

\noindent where $m_1 > \ldots > m_{2r} \geq 0$, 
$|\widehat{\mu}| = 
\sum_{j=1}^{2r}m_j$, 
$\alpha = 1$, for $m_{2r} \geq 1$, and 
$\alpha = \sqrt{2}$, for $m_{2r} = 0$.
An arrow on a product indicates the direction in which the value 
of the index of that product increases. 

\subsubsection{Remark} In KP theory, positively and negatively 
charged fermion modes translate to distinct horizontal and 
distinct vertical parts. These combine, in a standard way, to 
form partitions that are not necessarily strict \cite{blue-book}. 
In BKP theory, 
there are only neutral modes, which translate to one set of 
distinct parts, which form strict partitions. This is why 
only strict partitions appear in this work.

\subsection{A Heisenberg sub-algebra} We refer the reader to 
\cite{JM} for complete definitions of the infinite dimensional 
Lie algebra $B_\infty$, and its presentation in terms of bilinears 
in $\phi_m$. Here, all we need is the Heisenberg sub-algebra generated 
by $\lambda_m \in B_\infty$, where

\begin{equation}
\lambda_m = \frac{1}{2}\sum_{j \in \Z}(-)^{j+1}
\phi_j\phi_{-j-m}, \quad m \in \Z_{\mbox{\tiny{odd}}}
\label{bp}
\end{equation}

\noindent which satisfy the commutation relations

\begin{equation}
\left[ \lambda_m, \lambda_n \right] = \frac{m}{2}\delta_{m+n, 0}, 
\quad m,n \in \Z_{\mbox{\tiny{odd}}}
\label{bq}
\end{equation}

\begin{equation}
\left[ \lambda_m,\phi_n \right] = \phi_{n-m}, 
\quad m \in \Z_{\mbox{\tiny{odd}}},\ n \in \Z
\end{equation}

\subsection{Evolution operators} Writing 
$\Lambda_{\pm}(\mathbf{x}_{\mbox{\tiny{odd}}})$
$=$ 
$\sum_{m \in \pm \N_{\mbox{\tiny{odd}}}} x_m \lambda_m$, 
and
$\zeta_{\pm} (\mathbf{x}_{\mbox{\tiny{odd}}}, k)$ 
$=$
$\sum_{m \in \pm \N_{\mbox{\tiny{odd}}}} x_m k^m$, a standard 
computation shows that

\begin{equation}
\left[ \Lambda_{\pm}(\mathbf{x}_{\mbox{\tiny{odd}}}), 
\Phi(k) \right] = \zeta_{\pm} (\mathbf{x}_{\mbox{\tiny{odd}}}, k) 
\Phi(k)
\end{equation}

\noindent which implies 

\begin{equation}
e^{\Lambda_{\pm}(\mathbf{x}_{\mbox{\tiny{odd}}})}\Phi(k) 
e^{- \Lambda_{\pm}(\mathbf{x}_{\mbox{\tiny{odd}}})} =
\Phi(k) e^{\zeta_{\pm}(\mathbf{x}_{\mbox{\tiny{odd}}},k)}
\label{br}
\end{equation}

\noindent so that the operators 
$e^{\pm \Lambda_{\pm}(\mathbf{x}_{\mbox{\tiny{odd}}})}$ 
act as (forward and backward) evolution operators.

\subsection{A choice of parameters} Setting 
$x_m = \frac{2}{m}z^{-m}$, $m \in \Z_{\mbox{\tiny{odd}}}$, and 
writing 
$\Lambda_{\pm}(\mathbf{x}_{\mbox{\tiny{odd}}}) = \Lambda_{\pm}(z)$, 
and 
$\zeta_{\pm}(\mathbf{x}_{\mbox{\tiny{odd}}},k) = \zeta_{\pm}(z,k)$, 
we formally have

\begin{eqnarray}
\zeta_{+}(z,k) &=&
\phantom{-}
\sum_{m \in \N_{\mbox{\tiny{odd}}}}\frac{2}{m}
\ll \frac{k}{z} \rr^m 
=
\log{\ll \frac{z+k}{z-k} \rr} \label{bs} 
\\ \no 
\\ 
\zeta_{-}(z,k) &=& 
- 
\sum_{m \in \N_{\mbox{\tiny{odd}}}}\frac{2}{m}
\ll \frac{z}{k} \rr^m 
= 
\log{\ll \frac{k-z}{k+z} \rr} \no
\label{terms}
\end{eqnarray}

\subsection{Vertex operators}
Consider the vertex operators 

\begin{eqnarray}
\Gamma^{\phi}_{+}(z) 
= 
e^{\Lambda_{+}(z)} 
&=&
\exp{\ll \sum_{ m \in \N_{\mbox{\tiny{odd}}}}
\frac{2}{m}z^{-m} \lambda_m \rr} 
\label{bt} \\
\Gamma^{\phi}_{-}(z) 
=
e^{-\Lambda_{-}(z)} 
&=&
\exp{\ll \sum_{ m \in \N_{\mbox{\tiny{odd}}}}
\frac{2}{m}z^m\lambda_{-m} \rr}  
\label{bu}
\end{eqnarray}

Using Equations {\bf \ref{br}} and {\bf \ref{bs}}, 
expanding and equating powers of $k$, we obtain

\begin{eqnarray}
\Gamma^{\phi}_{+}(z) \phi_j \Gamma^{\phi}_{+}(-z) &=& 
\phi_j +2\sum_{n=1}^{\infty}\frac{1}{z^n}\phi_{j-n} \label{bv} \\ \no \\
\Gamma^{\phi}_{-}(-z) \phi_j \Gamma^{\phi}_{-}(z) &=& 
\phi_j +2\sum_{n=1}^{\infty} (-z)^n \phi_{j+n} 
\label{bw}
\end{eqnarray}

\subsection{A commutation relation} 
Commuting two vertex operators 

\begin{eqnarray*}
\Gamma^{\phi}_{+}(z) \Gamma^{\phi}_{-}(z') 
&=& 
e^{\Lambda_{+}(z)}e^{-\Lambda_{-}(z')} 
= e^{[\Lambda_{+}(z), -\Lambda_{-}(z')]}\ 
e^{-\Lambda_{-}(z')} e^{\Lambda_{+}(z)} \\ \\ 
&=&e^{[\Lambda_{+}(z),-\Lambda_{-}(z')]}\ 
\Gamma^{\phi}_{-}(z')\Gamma^{\phi}_{+}(z)
\end{eqnarray*}

\noindent and using 

\begin{eqnarray*}
\left[ \Lambda_{+}(z), -\Lambda_{-}(z') \right] &=&
\sum_{m \in \N_{\mbox{\tiny{odd}}}}
\sum_{n \in \N_{\mbox{\tiny{odd}}}}
\frac{4}{mn}z^{-m}(z')^{n}
[ \lambda_m, \lambda_{-n} ] 
\\ \\ 
&=& \sum_{m \in \N_{\mbox{\tiny{odd}}}}
\sum_{n \in \N_{\mbox{\tiny{odd}}}}
\frac{4}{mn}z^{-m}(z')^{n}\frac{m}{2}\delta_{m,n} \\ \\ &=& 
\sum_{m \in \N_{\mbox{\tiny{odd}}}}
\frac{2}{m} \ll \frac{z'}{z}\rr^m = 
-\log \ll \frac{z-z'}{z+z'} \rr
\end{eqnarray*}

\noindent we obtain the basic commutation relation 

\begin{equation}
\Gamma^{\phi}_{+}(z)\Gamma^{\phi}_{-}(z') = 
\ll
\frac{z+z'}{z-z'} 
\rr
\Gamma^{\phi}_{-}(z') \Gamma^{\phi}_{+}(z)
\label{ev-comm}
\end{equation}

\section{Interlacing strict partitions}\label{sp} 

In this section, we refer the reader to \cite{OR, ORV} for 
the definition of interlacing partitions.  We show how BKP 
vertex operators act on strict partitions to generate strict 
partitions that interlace with the initial ones. If 
$\widehat{\mu}$ and $\widehat{\nu}$ are interlacing, and 
$|\widehat{\mu}|$ $\geq$ $|\widehat{\nu}|$, where 
$|\widehat{\mu}|$ is the sum of the lengths of the parts of 
$\widehat{\mu}$, {\it etc},
we write $\widehat{\nu} \prec \widehat{\mu}$.

\begin{lemma}\label{lemma}
If $\widehat{\mu}$ and $\widehat{\nu}$ are strict partitions,
as in Definition {\bf \ref{strict}}, then 

\begin{equation}
\langle \widehat{\nu}|
\Gamma^{\phi}_{+}(z)
| \widehat{\mu} \rangle 
= 
\left\{
\begin{array}{ll}
2^{n(\widehat{\nu}|\widehat{\mu})}
z^{|\widehat{\nu}|-|\widehat{\mu}|}, & 
\quad\widehat{\nu} \prec \widehat{\mu} \ 
\mbox{and}
\ n(\widehat{\nu})=n(\widehat{\mu}) 
\bigskip 
\\ 
(-)^{n(\widehat{\mu})}
2^{n(\widehat{\nu}|\widehat{\mu})
+
\frac{1}{2}}z^{|\widehat{\nu}|-|\widehat{\mu}|}, 
&\quad 
\widehat{\nu} \prec \widehat{\mu} \ \mbox{and}\ n(\widehat{\nu})
=
n(\widehat{\mu})-1 \bigskip \\
0, & \quad \mbox{\it otherwise}
\end{array}
\right. 
\label{bz}
\end{equation}

\begin{equation}
\langle \widehat{\mu}|
\Gamma^{\phi}_{-}(z)
|\widehat{\nu}\rangle = 
\left\{
\begin{array}{ll}
2^{n(\widehat{\nu}|\widehat{\mu})}
z^{|\widehat{\mu}|-|\widehat{\nu}|}, & 
\quad
\widehat{\nu} \prec \widehat{\mu} \ \mbox{and}\ n(\widehat{\nu})=
n(\widehat{\mu}) \bigskip \\ 
(-)^{n(\widehat{\mu})}2^{n(\widehat{\nu}|\widehat{\mu})+\frac{1}{2}}
z^{|\widehat{\mu}|-|\widehat{\nu}|}, &\quad \widehat{\nu} \prec 
\widehat{\mu} \ \mbox{and}\ n(\widehat{\nu})=n(\widehat{\mu})-1 
\bigskip \\
0, & \quad \mbox{\it otherwise}
\end{array}
\right.
\label{ca}
\end{equation}

\noindent where 
$\{n(\widehat{\mu}), n(\widehat{\nu})\}$ 
is the number of non-zero parts in 
$\{\widehat{\mu}, \widehat{\nu}\}$, 
and $n(\widehat{\nu}|\widehat{\mu})$ 
is the number of non-zero parts in $\widehat{\nu}$ (the smaller
partition), that are not in $\widehat{\mu}$ (the larger
partition). It is important to notice that only
$n(\widehat{\nu}|\widehat{\mu})$ appears in both of the above 
equations, and not
$n(\widehat{\mu}|\widehat{\nu})$.
\end{lemma}

\noindent{\bf Proof.} Setting $m_{2r+1} = -1$, we have

\begin{eqnarray*}
\Gamma_{+}^\phi(z)| \widehat{\mu} \r 
&=&
\alpha (-)^r 
\Gamma_{+}^\phi(z) \rprod_{j=1}^{2r} \phi_{m_j} |0 \r 
= \alpha (-)^r \rprod_{j=1}^{2r} 
\ll
\Gamma_{+}^\phi(z) \phi_{m_j}
\Gamma_{+}^\phi(-z)
\rr
|0 \r 
\no \\ 
&=& 
\alpha (-)^r \rprod_{j=1}^{2r}
\ll
\phi_{m_j} + 2 \sum_{i=1}^{m_j-m_{j+1}-1}
\frac{1}{z^i} \phi_{m_j-i} + \frac{1}{z^{m_j-m_{j+1}}}
\phi_{m_{j+1}}
\rr
|0\r 
\no \\ 
\no \\ 
&=& 
\sum_{\substack{\widehat{\nu} \prec \widehat{\mu} \\ \\ 
n(\widehat{\nu}) = n(\widehat{\mu})}} 
2^{n(\widehat{\nu}| \widehat{\mu})}z^{|\widehat{\nu}|
-|\widehat{\mu}|}|
\widehat{\nu}\rangle+(-)^{n(\widehat{\mu})}\sqrt{2}
\sum_{\substack{\widehat{\nu} \prec \widehat{\mu} \\ \\ 
n(\widehat{\nu})=n(\widehat{\mu})-1}}
2^{n(\widehat{\nu}|\widehat{\mu})}
z^{|\widehat{\nu}|-|\widehat{\mu}|}|
\widehat{\nu}\rangle 
\no \\ 
\end{eqnarray*}

The proof of Equation {\bf \ref{ca}} goes along similar lines.

\subsection{Condition 1 on BKP plane partitions} We can now see 
the origin of condition {\bf 1}, stated above. As the vertex operators 
act on a diagonal slice to form the subsequent diagonal slice, 
Lemma {\bf \ref{lemma}} says that every time a new path starts, 
we pick up a factor of 2. When a path ends, there is no such 
contribution. This follows from the fact that 
$n(\widehat{\nu}|\widehat{\mu})$ appears in {\it both} equations 
in Lemma {\bf \ref{lemma}}, as mentioned above.

\section{Diagonally strict plane partitions}\label{dspp}

In this section, we show how interlacing strict partitions, 
stacked vertically, form diagonally strict plane partitions 
with $h$-paths that are 2-coloured and 1-column wide. 

\begin{de} Assuming that the highest column of a plane partition 
is at the north west corner, as in the example in Figure 
{\bf \ref{figure}}, a diagonally strict plane partition 
$\widehat{\pi}$ is a plane partition whose vertical slices, 
along all diagonals that run from north west to south east, 
are strict partitions. 
\end{de}

\subsection{Condition 2 on BKP plane partitions} The diagonal 
strictness condition does not allow any 4 height-$h$, $h > 0$, 
columns to be in a $2\times 2$ formation. Equivalently, every 
path is 1-column wide. Rather than give a formal, and definitely
tedious proof of this simple observation, we encourage the reader 
to verify it by experimenting on a few simple examples. 

\subsection{Generating BKP plane partitions} Following the choice 
of parameters used in \cite{nakatsu} and related papers, we consider 
the scalar product

\begin{eqnarray}
\widehat{S}(q) 
&=&
\langle 0| \ {\lprod_{j=1}^{\infty}
\Gamma^{\phi}_{+}\ll q^{\frac{-2j+1}{2}}\rr}\ 
{\rprod_{k=1}^{\infty}
\Gamma^{\phi}_{-}\ll q^{\frac{2k-1}{2}}\rr}|0
\rangle \label{cb} \\ \no \\ 
&=&
\sum_{\widehat{\mu}} 
\langle 0| \ {\lprod_{j=1}^{\infty}
\Gamma^{\phi}_{+} \ll q^{\frac{-2j+1}{2}}\rr}| \widehat{\mu}\rangle\langle
\widehat{\mu}|\ {\rprod_{k=1}^{\infty}
\Gamma^{\phi}_{-}\ll q^{\frac{2k-1}{2}}\rr}|0\rangle \label{cc}
\label{cd}
\end{eqnarray}

\noindent where $q$ is an indeterminate and $\sum_{\widehat{\mu}}$ 
indicates a sum over all strict partitions $\widehat{\mu}$. From
Lemma {\bf \ref{lemma}}, we know that 
$\Gamma^{\phi}_{+}(z)|\widehat{\mu}\rangle$ 
and 
$\langle\widehat{\mu}|\Gamma^{\phi}_{-}(z)$ generate all strict 
partitions $\widehat{\nu} \prec \widehat{\mu}$. As in \cite{nakatsu}, 
the arguments in the vertex operators are chosen such that the generated 
diagonally strict plane partitions are weighted by their volume. 
In addition, the vertex operators generate plane partitions with 
a multiplicity $2^{p( \widehat{ \pi})}$, where $p(\widehat{\pi})$ 
is the total number of $h$-paths in $\widehat{\pi}$. It follows 
that Equation {\bf \ref{cc}} receives a contribution of 

\begin{equation*}
{\lprod_{j=1}^{M} \langle \widehat{ \nu}_{-j}|
\Gamma^{\phi}_{+} \ll q^{\frac{-2j+1}{2}}\rr}|
\widehat{\nu}_{-j+1} \rangle\ 
{\rprod_{k=1}^{N} \langle \widehat{\nu}_{k-1}|
\Gamma^{\phi}_{-} \ll q^{\frac{2k-1}{2}}\rr}|
\widehat{\nu}_k\rangle 
 = 
2^{p(\widehat{\pi})} q^{|\widehat{\pi}|}
\end{equation*}

\noindent for each diagonally strict plane partition given by

\begin{equation*}
\widehat{\pi}=\left\{\emptyset=\widehat{\nu}_{-M} \prec \ldots \prec 
\widehat{\nu}_{-1} \prec \widehat{\nu}_0 \succ \widehat{\nu}_1 \ldots 
\succ \widehat{\nu}_N=\emptyset \right\} 
\end{equation*}

\noindent where $| \widehat{\pi} |$ is the volume of $\widehat{\pi}$.  
From that we conclude that

\begin{equation*}
\widehat{S}(q) =
\sum_{\widehat{\pi}}2^{p(\widehat{\pi})}q^{|\widehat{\pi}|}
\end{equation*}  

Applying the commutation relation Equation {\bf \ref{ev-comm}} 
repeatedly to Equation {\bf \ref{cb}}, one recovers a product 
form for the generating function $\widehat{S}(q)$ of plane 
partitions that satisfy the two conditions stated above:

\bigskip 
\begin{boxedminipage}[c]{12cm}

\begin{equation}
\widehat{S}(q) = \prod_{n=1}^{\infty}
\ll
\frac{1+q^n}{1-q^n}
\rr^n
\label{new}
\end{equation}

\end{boxedminipage}
\bigskip 

\section{Conclusion and remarks} In hindsight, Equation 
{\bf \ref{new}} is what we should have expected on the basis of 
the commutation relation of the vertex operators in Equation 
{\bf \ref{ev-comm}}. 
What is new is that Equation {\bf \ref{new}} counts plane partitions 
that do {\it not} form a subset of the plane partitions counted by 
MacMahon's well-known result, re-derived in \cite{OR}.

Clearly, it would be interesting to carry out a comprehensive study 
of plane partitions that are generated by various classes of free 
fermions. These would correspond to other integrable hierarchies, 
such as CKP, DKP, multicomponent hierarchies, such as $n$-KP, $n$-BKP, 
{\it etc}, and restricted versions of these, including KdV, {\it etc}. 
This is beyond the limited scope of this note, but we plan to report 
on them in further work.

An important question is whether the above result is relevant to
topological string theory. Since the result of \cite{ORV} relates 
KP theory, which is based on $A_{\infty}$ to a topological string 
that is dual to $U(N)$ Chern-Simons theory, in the limit 
$N \rightarrow \infty$, we naively expect that Equation {\bf \ref{new}}
is relevant to a topological string that is dual to $O(N)$ 
Chern-Simons theory, in the limit $N \rightarrow \infty$ \cite{SV}. 


\end{document}